\documentstyle[12pt]{article}
\newcommand{\beq}{\begin{eqnarray}}
\newcommand{\eeq}{\end{eqnarray}}

\newcommand{\ket}{\rangle}
\newcommand{\bra}{\langle}

\begin{document}

\begin{center}
{\bf MASS DEPENDENCE OF M3Y-TYPE INTERACTIONS AND THE EFFECTS OF 
TENSOR 
CORRELATIONS} \\                              
\vspace*{10mm}
 J. O. Fiase,  K. R. S. Devan \\
 \vspace{0.3cm}
Department of Physics, University of Botswana, P.B. 0022, Gaborone, 
Botswana\\
\vspace{0.3cm}
and\\
 \vspace{0.3cm}
A. Hosaka\\ 
 \vspace{0.3cm}
Research Center for Nuclear Physics (RCNP), Osaka University, Ibaraki 
567-0047, Japan\end{center}

\begin{abstract}
The mass dependence of the M3Y-type effective interactions and the 
effects of tensor correlations are examined. 
Two-body nuclear matrix elements are obtained by 
the lowest 
order constrained variational (LOCV) technique
with and without tensor correlations.  
We have found that the
tensor correlations are important especially in the 
triplet-even (TE) and tensor-even (TNE) channels in order to 
reproduce 
the G-matrix elements obtained previously.  
Then M3Y-type potentials for inelastic 
scattering are obtained by fitting our two body matrix elements to 
those 
of a sum of Yukawa functions for the mass numbers $A=24$, $A=40$ and 
$A=90$.

PACS number(s): 21.30. -x, 21.60.-n, 24.10. -i   

\end{abstract}

\section{Introduction}

The study of inelastic reactions from fundamental 
nucleon-nucleon interaction has been a major topic for nearly 30 
years now following the pioneering G-matrix work of Bertsch and 
collaborators [1]. In particular their M3Y interaction [1] and its 
density-dependent versions[2] have shown that one can predict quite 
unambiguously inelastic scattering from fundamental nucleon nucleon 
interaction. Because of their semi-microscopic approach, there has 
been a lot of interest in these interactions by many researchers 
[3,4].

Recently [5] we have produced a similarly motivated potential for 
$A=16$ nuclei which was based on the  lowest 
order constrained variational (LOCV) approach.   
We have 
found that our interaction was very similar in the singlet-even, 
triplet-even, tensor-even and spin-orbit  odd channels with those 
of the M3Y interaction but differed significantly in the 
singlet-odd, triplet-odd and spin-orbit even channels. 
It 
thus appears that the issue of finding precise strengths of 
interactions in these channels has not yet been settled and still 
requires some further attention. The validity of using the M3Y 
interaction for inelastic scattering has been tested in many 
applications [6] and  we believe that it is possible to improve our 
understanding of  inelastic scattering from realistic forces  if all 
the correct ingredients such as finding the precise strengths of the 
interaction in all angular momentum channels and its mass dependence 
are  considered in a systematic way. \\
Finding these ingredients may not 
be very easy and may require going beyond the usual non-relativistic 
two-body approach of nucleon-nucleon forces to include for example, 
the three-body forces and isobar degrees of freedom. Besides these 
considerations we must check carefully our models since these 
quantities may also be  model-dependent. 
For example, in the G-matrix 
approach, it was found that the triplet-even (TE)  matrix elements 
used in constructing the inelastic potentials 
depended sensitively on the starting energies [1]. This is a problem 
of model-dependence whereas in our model, these same quantities 
depend sensitively on the strength of the tensor correlations. 
Without tensor correlations we shall show in this paper that we have 
no agreement in these channels with the G-matrix approach.

Another important consideration is the mass dependence. It is useful 
to consider the mass dependence of such an interaction in a 
systematic way while using it on all regions of nuclei. For example, 
while the M3Y interaction has been very successful in explaining 
inelastic scattering from fundamental forces, it was constructed for 
the $A = 16$ nuclei. One is not sure how it varies with the mass 
number, $A$. This last point on mass dependence has been emphasized in 
the classic work of Wildenthal [7] who found that it was impossible 
to find a mass - independent two-body effective interaction that 
could explain all sd -shell nuclei.

The purposes of the present work are two-fold: 
(i) to extend the work of ref. 
[5] which was designed for the $A= 16$ system by including the mass 
dependence of our potential model 
for inelastic scattering and (ii) to study the effects of tensor 
correlations which affects particularly  the TE and TNE channels. 
Other ingredients such as the inclusion of  density dependence, 
addition of three-body forces and isobar degrees of freedom etc could 
also be considered but we shall leave them out from this paper and 
consider our approach a first approximation.

The paper is organized as follows: In section 2 we give a brief 
summary of the method used. In section 3, we define the $NN$
interaction to be constructed. In section 4 we discuss our findings in the present work. The final section  is devoted to the conclusion of the 
paper.

\section{Formalism}
In this section we briefly review the LOCV prescription discussed by 
Irvine and collaborators [8,9] for obtaining an effective two-body 
interaction which is adopted for the present work. In this 
approach 
the approximation to a non-relativistic nucleon fluid interacting 
through a two-body potential is first approximated by the Hamiltonian 
\begin{equation}
H_{0}=\sum_{i}-\frac{\hbar^2}{2m} 
\bigtriangledown_{i}^2+\sum_{i>j}V_{ij},
\end{equation} 
where $V_{ij}$ is the two-body potential. In the LOCV approach the 
translationally invariant component of the trial wavefunction is 
defined as\\
\begin{equation}
\Psi_{T}= U F \Phi,
\end{equation}  
where $U$ is a unitary operator which transforms the system to the 
centre of mass rest-frame, leaving us with only intrinsic quantities 
such that we do not have to worry any further about the spurious 
centre of mass 
motion. 
In eq. (2), $F$ is a symmetric product of two-body correlation 
functions defined as [9]:
\begin{equation}
F=\prod_{i>j}f_{2}(ij), 
\end{equation}  
designed  to accommodate the effect of the strong repulsion of the 
nucleon-nucleon interaction, and $\Phi$ is a multi-dimensional 
product of two-body wave functions. The explicit form of $f_{2}(ij)$ is given in eqs. (10) -(13) below.

Our task is to evaluate the expectation value of the Hamiltonian 
given by
\begin{equation}
E=\frac{\bra \Psi_{T}\mid H_{0}\mid\Psi_{T} \ket}
{\bra \Psi_{T}\mid\Psi_{T} \ket}
\end{equation}
with the desired accuracy. 
Due to its multi-dimensional nature, eq. 
(4) is difficult to calculate and we must make approximations. 
The cluster expansion technique is the approximating technique we 
adopt. Here
we divide the system into clusters. 
We evaluate the 
energy of each cluster starting with the two-body clusters and then 
sum over all the clusters to obtain the total energy:
\begin{equation}
E= E_{2}+E_{3} +E_{4}+...\, \, \, \,\,.
\end{equation}
To lowest order we minimize the two-body energy term $E_{2}$ with 
respect to 
the functional variations of the two-body correlation functions such 
that only two-body cluster terms are important, whereas 
contributions of $E_{3}$ and higher order clusters are made 
small. 
This is achieved if the convergence parameter $k$, defined by the 
equation below is small [10];
\begin{equation}           
k= \frac{3}{r_{0}^{3}}\int
\bra (1 - f_{2}(ij))^2  \ket
{r_{ij}^{2}{dr_{ij}} \ll 1} \, ,                                                                                          
\end{equation} 
where $r_{0}$ is the mean nearest neighbour distance, defined from the 
nuclear radius $R$ such that:
\begin{equation} 
R = r_{0}A^{1/3},
\end{equation}
while $<...>$ signifies channel average, including the tensor operator.
This approximation implies that we have accepted to work only with 
two-body clusters. Thus our $E_{2}$ takes the form  [10]:          
\begin{equation}
E_{2}= \bra \Phi\mid\sum_{i>j}( f_{2}(ij)
( P_{ij}^{2}/M + V_{ij}) 
f_{2}(ij))\mid\Phi 
\ket                                                                
\end{equation}                 
where $\vec{P}_{ij} = \frac{1}{\sqrt{2}}(\vec{p}_{i} - \vec{p}_{j})$ 
is the relative momentum of the two-particle system; 
$M \sim m_{N}A$ is 
the total mass of the nucleus and $V_{ij}$ is taken to be the Reid 
[11] soft - core potential.
  
Since the Reid [11] potential has the form:
\begin{equation}
V_{ij}=\sum_{\lambda}V_{ij}^{\lambda},
\end {equation}
where in different reaction channels $\lambda$, we have the 
central, spin-orbit and tensor components.  
For each channel we introduce 
two-body correlation functions, 
\begin{equation}
f_{2}(ij)=\sum_{\lambda}f_{ij}^{\lambda}
\end {equation}
where
\begin{equation}
f_{ij}^{\lambda} =
f_{c}^{\lambda}(r_{ij}) 
+ f_{LS}^{\lambda}(r_{ij})\vec{L} \cdot \vec{S} 
+ f_{T}^{\lambda}(r_{ij})S_{ij}\,, 
\end {equation}
with the tensor operator
\begin{equation}
    S_{ij}= 3 \vec \sigma_{i} \cdot \hat r_{ij} \
    \vec \sigma_{j} \cdot \hat r_{ij} 
    - \vec \sigma_{i} \cdot \vec \sigma_{j} \, .
\end{equation}

In studies regarding nuclear matter and finite nuclei [9], the 
two-body correlation functions were found to have three features  
which included the 'wound' induced in the two-body wave function by 
the repulsive core of the N-N interaction, the tensor correlations 
especially in the $^{3}S_{1} - ^{3}D_{1}$ channel and the meson 
exchange correction. It was found that the most important  feature of 
these was the tensor correlations. Therefore the two-body correlation 
functions of eq. (10) have been parametrized in the form [10,12,13]:
\begin{eqnarray}
r_{ij} < r_{c}  &:& \;  
f^{\lambda}_{ij} = 0\, ,  \nonumber\\
r_{ij}\geq{r_{c}} &:& \; 
f^{\lambda}_{ij} =  (1 - 
e^{-\beta{(r_{ij}-r_{c})}^{2}})({1+\alpha^{\lambda}(A){S_{ij}}})\, , 
\end{eqnarray}                                                            
where $r_{c} = 0.25fm$ and $\beta = 25fm^{-2}$. The parameter, 
$\alpha^{\lambda}(A)$ represents the strength of the tensor 
correlations 
and is non-zero only in the $^{3}S_{1}$ - $^{3}D_{1}$ and $^{3}P_{2}$ 
- $^{3}F_{2}$ channels.
 
It should be noted that this form of correlation functions is clearly an over simplification. In reality correlation functions are highly density-dependent and are different in different tensor channels. Moreover, they are usually obtained from nuclear matter calculations. We have retained this form because of its previous success [9,10,12,13] and because we wish to study the effects of tensor correlations on our model calculations where we have devised a method of switching on and off the tensor correlations.

In order to evaluate matrix elements we work in the harmonic 
oscillator basis. For our purpose we have picked out only the 
effective 
potential energy terms from eq. (8) i.e.,
\begin{equation}
E_{2}^{'}= \bra \Phi\mid\sum_{i>j} f_{2}(ij) V_{ij} 
f_{2}(ij)\mid\Phi 
\ket                                                            
\end{equation}   
Here the only parameters appearing in our calculations are; the 
oscillator 
size parameter and the strength of the tensor correlations, 
$\alpha^{\lambda}(A)$. We shall later present our results with 
various 
combinations of these parameters with and without tensor 
correlations. Furthermore  following the procedure of Berstch and 
collaborators [1] we have separated the relative potential two-body 
matrix elements into those of 
various channels. 
These are the singlet-even 
(SE) and singlet-odd (SO)
channels denoted by $^{1}S_{0}$  and $^{1}P_{1}$ respectively. The 
triplet-even (TE) and the tensor-even (TNE)components were picked 
from the coupled $^{3}S_{1}$ - $^{3}D_{1}$ channels. We next define 
the triplet-odd (TO), tensor-odd (TNO) and the two components of the 
spin-orbit force ignoring the quadratic term as [14]:
\begin{eqnarray}
V(TO)  = V(^3P_0)+2V(LSO) +4V(TNO)\nonumber\\
V(TNO) = - \frac{5}{72}[ 2V(^3P_0)-3V( ^3P_1)+V(^3P_2)] 
\nonumber\\                                            
V(LSO) = - \frac{1}{12}[2V(^3P_0)+3V(^3P_1)-5V(^3P_2)] \nonumber\\
V(LSE) =  \frac{1}{3}[V(TE)-2V(TNE)-V(^3D_1)]\,.
\end{eqnarray}

\section{The $NN$ Effective Interaction}

In this section we define an effective interaction which is suited
to calculations of inelastic scattering. 
We follow the procedure described by Bertsch and collaborators [1] in 
which the two-body matrix 
elements are fitted to those of the sum of Yukawa functions with 
different ranges.   
The potential is 
divided into the central (c), spin-orbit (ls) and tensor (t) 
components as follows: 
\begin{eqnarray} 
V_{c}  &=& \sum_{k}d_{k} 
\frac{e^{-(r_{ij}/R_{k})}}{(r_{ij}/R_{k})},\nonumber\\
V_{ls}  &=& \sum_{k}d_{k} \frac{e^{-(r_{ij}/R_{k})}}{(r_{ij}/R_{k})} 
\vec{L}.\vec{S},\nonumber\\      
\label{effV}
V_{t}  &=& \sum_{k}d_{k} 
\frac{e^{-(r_{ij}/R_{k})}}{(r_{ij}/R_{k})}r_{ij}^2 S_{ij},
\end{eqnarray}
where the $ d_{k} $ are the strengths of the interaction which are 
determined by fitting the oscillator matrix elements of eq. (16) 
to our two-body matrix elements of eq. (14). 
The  ranges 
$k \leq 4 $ are chosen to be  0.25, 0.4, 0.7 and 1.414 fm, which are 
motivated by the one-boson exchanges. 
Apparently the longest range 1.414 fm corresponds to 
the one-pion exchange, 
while the shorter ranges correspond to heavier mesons such as 
$\sigma$, $\rho$  and $\omega$ mesons.

\section{ Results}  

The objective of our present paper is to extend the work of ref. [5] 
in which we defined an effective interaction for inelastic scattering 
for the $A= 16$ system. In the present paper we incorporate the mass 
dependence as well as investigate the effects of tensor correlations 
on our earlier calculations [5].

In Table 1 we present the relative two - body matrix elements 
of the effective interaction for the $A=24$ system for the SE, SO, TE, 
TO, TNE, TNO, LSO and LSE channels. The matrix elements were 
calculated in a harmonic oscillator basis with 
$\hbar\omega=45A^{-1/3}-25A^{-2/3}$MeV so that $\hbar\omega$ takes on
the value of 12.7 MeV in this case. Based on our observations in 
[8,12] for the $sd$ shell and earlier calculations for the $p$ shell 
[10], it was found that $\alpha^{\lambda}(A)$ is a monotonic 
decreasing function of the mass number, A and ranges from 0.09 at the 
beginning of the $p$ shell to about 0.07 for the upper end of the $sd$ 
shell. When extrapolated to the $A=90$ region, it takes the value of 
about 0.06. Using this information, the calculation is first 
performed with tensor correlations switched on, i.e. by setting 
$\alpha^{\lambda}(A)= 0.075$ appropriate for the A = 24 system. 
However, when $\alpha^{\lambda}(A)=0.0$, we have only central 
correlations.  
We should observe that the 
tensor correlations are effective only in the coupled  channels such 
as  $^{3}S_{1}$ - $^{3}D_{1}$ and $^{3}P_{2}$ - 
$^{3}F_{2}$, while the  SE and SO channels are not subject to the 
tensor correlations.  

In Table 2 we repeat the same procedure as in Table 1 for the $A = 40$ 
system but this time we have used $\alpha^{\lambda}(A)= 0.070$ which 
we found most appropriate for this system with $\hbar\omega = 11.0 
MeV$ while for the $A = 90$ system  presented in Table 3 we have used 
$\alpha^{\lambda}(A)= 0.060$  and $\hbar\omega = 8.8 MeV$ based on 
our observations above.

\subsection{Central Even Channels}

From  Table 1,  we notice an 
interesting
result for the TE 
channel when the strength of the tensor correlations 
$\alpha^{\lambda}(A)$ is 
switched off. With $\alpha^{\lambda}(A)$ turned off the TE matrix 
elements are large 
and positive in our calculation. However, with $\alpha^{\lambda}(A)$ 
turned on they become large and negative. 
For an $S$-wave ($^3S_{1}$) channel, this attraction is caused by the 
second order (two pion exchange) contribution with the $D$-wave 
($^3D_{1}$) mixing.  
The role of the pion in this channel has long been appreciated.  
Recently, it is emphasized even more in exact calculations of few 
body systems [15] and in the relativistic mean field theory [16].   
The TO, TNO and LSO channels are also 
affected by the tensor correlations through the $^{3}P_{2}$  
state. The LSE channel is affected by the tensor correlations through 
the $^{3}D_{1}$ state, however, these effects are not as dramatic as 
those  observed in the TE channel case.  The SE and SO channels are 
not affected by the tensor correlations as indicated in the table. 
For the $A=24$ system  the  G-matrix data was unavailable with which to 
compare our results. 

In Table 2 we notice the similar results 
for the TE channel when the strength of the tensor correlations 
$\alpha^{\lambda}(A)$ is switched on or off. 
Here also our calculated TE 
matrix elements are large and positive whereas with 
$\alpha^{\lambda}(A)$ turned on they are large and negative. These 
should be compared with the G-matrix  results  of Hosaka et al. [17] 
for the $A=40$ system. Clearly, with $\alpha^{\lambda}(A)$
switched on to take the value of 0.075, we obtained a 
near perfect agreement for our calculated TE matrix elements with the 
G-matrix result of Hosaka et al.[17] 
where the difference between the two calculational procedures for 
each ($n'$,n) values is less than 12 percent. 
Also as shown in  Table 2 except for the ($n'$,n)=(2,2) quantum 
numbers, our calculated  SE matrix elements are found to agree with 
the G-matrix results of Hosaka et al.[17] to within 25 percent.

In Table 3 for the $A=90$ system we notice a similar trend as in Tables 
1 and 2. The same observation for the TE channel is inferred in Table 
3. Here we have used small node quantum numbers to compare our 
results with the G-matrix results. We notice again that the TE 
comparison with the G-matrix results is impressive when 
$\alpha^{\lambda}(A)$  is switched on.
This analysis supports the importance of tensor correlations in 
effective interaction theories that use correlated basis functions. 
Therefore in subsequent analysis we shall only be concerned with 
matrix elements when  
$\alpha^{\lambda}(A)$  is non-zero.

\subsection{Central Odd Channels}
As can be seen from Table 2 for the $A=40$ system, our calculated  SO 
matrix elements are 
similar to the G-matrix for the lower node quantum numbers but they 
become substantially different for higher node quantum numbers where 
for example, for the ($n', n$)=(2,2), they differ by over 80 percent. 
This difference between our calculated matrix elements and the 
G-matrix results in the odd angular momentum channels persists and  
can be clearly seen in the TO channel where our calculated matrix 
elements are positive while the G-matrix results are small and 
negative.
We observe a similar trend to Table 2 in Table 3 for the $A=90$ 
system.  

\subsection{Tensor Channels}
From Table 2 we see that the TNE channels are best represented. In 
these channel, our 
calculated matrix elements do not differ from the G-matrix 
calculations by more than 10 percent except for the ($n',n$)=(0,0) 
and (2,2) channels. 
The TNO matrix elements are also very similar to the 
G-matrix results with our calculated values slightly smaller than the 
G-matrix values. In Table 3 we notice a 
good agreement 
of our calculated TNE matrix elements with their G-matrix 
counterparts for the small node quantum numbers.  Our calculated TNO 
matrix elements are also seen to be in good agreement with the 
G-matrix results.

\subsection{Spin-Orbit Channels}
In these channels, we  notice as indicated in Tables 2 and 3 that our 
calculated LSO matrix elements are in close agreement with the 
G-matrix values.   
However, the LSE matrix elements are different 
from their G-matrix counterparts, although absolute values of the 
matrix elements are small.   
Here, the G-matrix results are 
negative while our calculated matrix elements are all 
positive.\\
It could be seen that in all channels, the matrix elements are 
decreasing in absolute values with increasing mass number, $A$. 

\subsection{Strength of Interactions}

We have performed a least squares fit of our  calculated two - 
body matrix elements in the various channels to those of a sum of 
Yukawa 
potentials for the central, tensor and spin-orbit forms as given in 
eq. (16). These forms are consistent with potentials for 
inelastic scattering. The selected ranges of 0.25, 0.4, 0.7 and 1.414 
fm were 
theoretically motivated and were chosen so as to ensure the OPEP 
tails in the relevant  channels as well as the short-range part which 
account for the exchange processes. Our interaction strengths are 
presented in Tables 4, 5 and 6 for  $A=24$, 40 and 90 systems respectively.
It should be noted that  the oscillator matrix elements of the radial 
components of eq. (16) decrease with increasing $A$, i.e., with 
decreasing $\hbar\omega$. 
This is because wave functions extend more as $A$ is increased, while 
the range of the interaction is finite and fixed.  
Our observation here is that the major part of the mass dependence of 
matrix elements is due to the change in the wave function.  
In order to further study mass dependence of the interaction 
strength, we need 
more detailed information of the effective interaction.  

%

\section{Conclusions}

The two calculational procedures presented here raise several 
interesting observations: 
The first observation is that although there is some model 
dependence 
in the calculation of the matrix elements in the TE channels
(such as sensitively  on the starting energies used 
in the calculation [1]), 
reasonable theoretical methods predict attractive matrix 
elements through a 
significant contribution from the tensor correlations. 
This indicates the importance of the tensor force in the 
nucleon-nucleon interaction dominated by the one pion exchange.  

Secondly, we notice that while the matrix 
elements calculated in our present method are in excellent  agreement 
with the G-matrix results of, for instance,  
Hosaka et al. [17], in most angular 
momentum channels, there is a problem in fixing 
the interaction strengths especially in the odd angular 
momentum channels. 
This problem was also found in ref.[5] for the 
$A= 16$ system. This can be seen in the SO, TO and the LSE channels. 
In fact, Bertsch and collaborators [1] had to use the set of Elliott 
[18] matrix elements in  the odd angular momentum  channels to get a 
good fit for their M3Y interaction. Hosaka et al.[17] have traced 
this 
problem to the fact that because the harmonic oscillator wave 
function vanishes near the origin for the odd forces, their exist an 
ambiguity in fitting the G-matrix elements in the odd angular 
momentum channels.  

The calculated matrix elements show an overall decrease in magnitude 
as we go from $A= 24$ to $A=90$ nuclei through the $A$-dependence of 
$\hbar\omega$ and with the moderate mass dependence of the strength of 
the tensor 
correlations $\alpha^{\lambda}(A)$. We have calculated the strengths 
of our interaction for $A=24$, 40  and $A=90$ nuclei as reported here. 
Since these depend on the mass 
number $A$, It will be interesting to see if this mass dependence may 
affect calculations based on this interaction. 

Thus, apart from the necessary inclusion of the density dependence 
and other effects in 
effective two-body interactions, one should not ignore 
perhaps the mass dependence and the model dependence in such 
calculations  in a systematic way.
\vspace*{0.5cm}

{\bf ACKNOWLEDGMENTS}

One of us (J.O.F) wishes to thank Professor J. M. Irvine who introduced him to the theory of  effective two-body interactions.
 
\newpage

\begin{table}[tbp]
   \centering
   \caption{\small Calculated relative matrix elements for $A = 24$ 
using 
$\hbar \omega = 12.7 MeV$. 
The first entry for a column is the result 
of the present calculation with $\alpha = 0.075$.  
The second entry in parentheses is the result 
of the present calculation with $\alpha = 0.0$. 
There  were no G-matrix calculations for comparison for this system. 
Notice that the SE and SO channels are not affected by the tensor 
correlations. }
\vspace*{0.5cm}
\begin{tabular}{rlll | rlll}
\hline
SE(S/S)   &  $n = 0$  &  1  &  2 &
    TE(S/S)     & $n = 0$    &       1    &  2   \\  
\hline
$n'= 0$  & $-7.02$   &  $-6.39$   &  $-5.33$    &          
$n'= 0$  &  $-10.37$ & $-8.34$  & $-5.99$  \\      
& & & &
&  $(3.69)$   & $(5.50)$    & $(6.92)$    \\
 1       &           &  $-6.29$   &  $ -5.42$   &  
   1    &            & $-7.71$    & $-5.84$    \\          
         &           &  &    &
        &            & $(7.34)$  & $(8.82)$  \\        
2        &           &            &  $- 4.87$   &
   2    &            &            & $-4.75$   \\
 & &            &            
& & & &$(10.17)$  \\
         
\hline
 
\hline
SO(P/P)   & $n=0$    &     1     &    2        & 
TO(P/P)   & $n=0$    &     1     &    2        \\
\hline
$n'= 0$   & $1.84$   & $2.78$    & $3.56$       &              
$n'= 0$   &  0.65    & 0.78      &  0.87       \\
 &          &     &       &
           & $(0.11)$  & $(0.18)$ & $(0.29)$   \\      
  1       &          &  4.28     & 5.29        &        
  1       &          &  1.12     & 1.31        \\          
          &          &     &       &
          &          & $(0.34)$  & $(0.47)$    \\
  2       &          &           &  6.55        &
  2       &          &           &  1.60       \\
          &          &           &       &
          &          &           &   (0.66)    \\
\hline

\hline
TNE(S/D)   & $n =0$  &       1     &        2  & 
TNO(P/P)   & $n =0$  &       1     &        2  \\
\hline
$n'= 0$ & $-4.42$    & $-6.36$     & $-7.64$   &               
$n'= 0$ &   0.50     &  0.46       & 0.36      \\                
        & $(-5.35)$  & $(-7.68)$   & $(-9.23)$ &
        &  (0.57)    &   (0.53)    &  (0.43)   \\
        
  1     & $-2.49$    & $ -4.70$    & $-6.45$   &       
  1     &            &  0.52       &   0.43  \\
        & $(-2.99)$  & $ (-5.70)$  & $(-7.89)$ &
        &            &   (0.61)    &   (0.53)  \\
  2     & $-1.51$    & $-3.05$     & $-4.73$   &
  2     &            &             &  0.39     \\
        & $(-1.84)$  & $(-3.79)$   & $(-5.92)$ &
        &            &             &  (0.51)   \\
        
\hline
\hline
 LSE(D/D)   &   $n = 0$   &    1       &    2      & 
 LSO(P/P)   &   $n = 0$   &    1       &    2      \\
\hline
$n'= 0$     &    0.22     &   0.25    & 0.24     &               
$n'= 0$     &   $-0.22$   & $-0.51$    & $-0.75$   \\                
            & (0.05)    & $(0.07)$  & $(0.08)$ &
            & $(-0.63)$   & $(-0.95)$  & $(-1.18)$ \\
    1       &             &   0.34     &   0.37    &
    1       &             &  $-0.83$   & $-1.12$    \\          
            &             &  $(0.11)$ & $(0.13)$ &
            &             &  $(-1.40)$ & $(-1.74)$  \\
    2       &             &            &   0.44     &
    2       &             &            & $- 1.44$    \\
            &             &            & $(0.16)$  &
            &             &            & $(-2.15)$  \\
   
\hline
\end{tabular}

\end{table}

\newpage

\begin{table}[tbp]
   \centering
   \caption{\small Calculated relative matrix elements for $A = 40$ 
using 
$\hbar \omega = 11.0 MeV$. 
The first entry for a column is the result 
of the present calculation with $\alpha = 0.070$. 
The second entry is the G-matrix calculation with the 
Paris potential.
The third entry in parentheses is the result 
of the present calculation with $\alpha = 0.0$.  }
\vspace*{0.5cm}
\begin{tabular}{rlll | rlll}
\hline
SE & & & & TE & & & \\
(S/S)   &  $n = 0$  &  1  &  2 &
(S/S)     & $n = 0$    &       1    &  2   \\  
\hline
$n'= 0$  & $-5.96$   &  $-5.62$   &  $-4.87$    &          
$n'= 0$  &  $-8.26$ & $-6.88$  & $-5.17$  \\                    
         & $-5.2187$   &  $-4.5601$   &  $-3.6117$    &
        &  $-8.040$   & $-7.4460$    & $-6.3224$    \\
&        &            &           &
          &  $(2.85)$   & $(4.24)$    & $(5.35)$ \\

 1       &           &  $-5.67$   &  $ -5.05$   &  
   1    &            & $-6.47$    & $-5.09$    \\          
         &           &  $-4.2845$ &  $-3.4698$  &
        &            & $-7.3175$  & $-6.3477$  \\
&        &            &           &
          &           & $(5.69)$    & $(6.87)$ \\        
2        &           &            &  $- 4.67$   &
   2    &            &            & $-4.25$   \\
         &           &            &  $ -2.8761$ &
        &            &            & $-5.6668$  \\
&        &            &           &
          &           &           & $(7.99)$ \\
\hline
SO & & & & TO & & & \\
(P/P)   & $n=0$    &     1     &    2        & 
(P/P)   & $n=0$    &     1     &    2        \\
\hline
$n'= 0$   & $1.31$   & $1.96$    & $2.55$       &              
$n'= 0$   &  0.48    & 0.56      &  0.64       \\
          & $1.7260$ & $1.7184$  & $1.5952$    &
          & $-0.0483$ & $-0.0893$  & $-0.0967$    \\
&        &            &           &
          &  $(0.096)$         & ($0.14)$    & $(0.18)$ \\  
  1       &          &  3.08     & 3.87        &        
  1       &          &  0.81     & 0.93        \\          
          &          & 2.2136    & 2.2640       &
          &          & $-0.0970$  & $-0.0939$    \\
&        &            &           &
          &          & $(0.23)$    & $(0.33)$ \\
  2       &          &           &  4.86        &
  2       &          &           &  1.14       \\
          &          &           &   2.5781    &
          &          &           &   -0.0701    \\
&        &            &           &
          &          &            & $(0.46)$\\
\hline
TNE & & & & TNO & & & \\
(S/D)   & $n =0$  &       1     &        2  & 
(P/P)   & $n =0$  &       1     &        2  \\
\hline
$n'= 0$ & $-3.49$    & $-5.08$     & $-6.17$   &               
$n'= 0$ &   0.41     &  0.39       & 0.33      \\                
        & $-3.8897$  & $-5.5050$   & $-6.4728$ &
        &  0.5374    &   0.5326    & $ 0.4738$   \\
 & $(-4.19)$       &  $(-6.08)$ & $(-7.37)$&
 & $(0.46)$         &  $(0.45)$        & $(0.38)$ \\
       
  1     & $-2.06$    & $ -3.84$    & $-5.27$   &       
  1     &            &  0.46       &     0.41  \\
        & $-2.0937$  & $-3.8428$  & $-5.2369$ &
        &            &   0.6556    &   0.6460  \\
        & $(-2.45)$  &  $(-4.59)$ & $(-6.34)$&
        &         &  $ (0.53)$        & $(0.49)$ \\

  2     & $-1.30$    & $-2.58$     & $-3.95$   &
  2     &            &             &  0.40     \\
        & $-1.1172$  & $-2.3980$   & $-3.6559$ &
        &            &             &  0.7073   \\
   & $(-1.55)$       &  $(-3.13)$ & $(-4.83)$&
        &         &               & $(0.49)$ \\
\hline
 LSE & & & & LSO & & & \\
 (D/D)   &   $n = 0$   &    1       &    2      & 
 (P/P)   &   $n = 0$   &    1       &    2      \\
\hline
$n'= 0$     &    0.15     &   0.19    & 0.18     &               
$n'= 0$     &   $-0.15$   & $-0.35$    & $-0.53$   \\                
            & -0.0059     & $-0.0090$  & $-0.0091$ &
            & $-0.3261$   & $-0.4751$  & $-0.5801$ \\
	&  (0.03)    &     (0.05)      & (0.06)       &
          &  $ (-0.44)$   & $(-0.68)$  & $(-0.86)$ \\ 
    1       &             &   0.25     &   0.27    &
    1       &             &  $-0.59$   & $-0.82$    \\          
            &             &  $-0.0278$ & $-0.0283$ &
            &             &  $-0.6969$ & $-0.8546$  \\
		&		  &  $(0.08)  $ & $(0.09)$&
            &             &  $-1.02    $ & $-1.28$   \\
    2       &             &            &   0.33     &
    2       &             &            & $- 1.07$    \\
            &             &            & $-0.0553$  &
            &             &            & $-1.0522$\\
		&		  &            & $(0.12)$ &
            &             &            & $(-1.59)$ \\
 	
\hline
\end{tabular}

\end{table}
\newpage

\begin{table}[tbp]
   \centering
   \caption{\small Calculated relative matrix elements for $A = 90$ 
using 
$\hbar \omega = 8.8 MeV$. 
The first entry for a column is the result 
of the present calculation with $\alpha = 0.060$. 
The second entry is the G-matrix calculation with the 
Paris potential.
The third entry with parentheses is the result 
of the present calculation with $\alpha = 0.0$. }
\vspace*{0.5cm}
\begin{tabular}{rlll | rlll}
\hline
SE & & & & TE & & & \\
(S/S)   &  $n = 0$  &  1  &  2 &
(S/S)     & $n = 0$    &       1    &  2   \\  
\hline
$n'= 0$  & $-4.58$   &  $-4.51$   &  $-4.09$    &          
$n'= 0$  &  $-5.43$ & $-4.69$  & $-3.68$  \\                    
         & $-4.1471$   &  $-3.8435$   &  $-3.2648$    &
        &  $-6.4335$   & $-6.2354$    & $-5.5745$    \\
&        &            &           &
          &  $(1.89)$ & $(2.82)$  & $(3.57)$ \\

 1       &           &  $-4.69$   &  $ -4.37$   &  
   1    &            & $-4.48$    & $-3.65$    \\          
         &           &  $-3.7805$ &  $-3.2868$  &
        &            & $-6.3463$  & $-5.7768$  \\
&        &            &           &
          &           & $(3.80)$  & $(4.63)$ \\        
2        &           &            &  $- 4.18$   &
   2    &            &            & $-3.09$   \\
         &           &            &  $ -2.9243$ &
        &            &            & $(-5.3810)$  \\
&        &            &           &
          &           &           & $(5.44)$ \\
\hline
SO & & & & TO & & & \\
(P/P)   & $n=0$    &     1     &    2        & 
(P/P)   & $n=0$    &     1     &    2        \\
\hline
$n'= 0$   & $0.77$   & $1.13$    & $1.49$       &              
$n'= 0$   &  0.27    & 0.33      &  0.35       \\
          & $1.2020$ & $1.2299$  & $1.1562$    &
          & $-0.0256$ & $-0.0591$  & $-0.0745$    \\
&        &            &           &
          &  $(0.04)$     & $(0.06)$  & $(0.08)$ \\  
  1       &          &  1.82     & 2.33        &        
  1       &          &  0.46     & 0.53        \\          
          &          & 1.5768    & 1.6314       &
          &          & $-0.0751$  & $-0.0843$    \\
&        &            &           &
          &          & $(0.14)$  & $(0.18)$ \\
  2       &          &           &  2.99        &
  2       &          &           &  0.66       \\
          &          &           &   1.8566    &
          &          &           &   -0.0813    \\
&        &            &           &
          &          &            & $(0.26)$\\
\hline
TNE & & & & TNO & & & \\
(S/D)   & $n =0$  &       1     &        2  & 
(P/P)   & $n =0$  &       1     &        2  \\
\hline
$n'= 0$ & $-2.41$    & $-3.58$     & $-4.40$   &               
$n'= 0$ &   0.29     &  0.30       & 0.27      \\                
        & $-2.7015$  & $-3.9266$   & $-4.7104$ &
        &  0.3756    &   0.3895    & $ 0.3591$   \\
 & $(-2.84)$     &  $(-4.20)$& $(-5.16)$&
 & $(0.32)$      &  $(0.33)$        & $(0.30)$ \\
       
  1     & $-1.52$    & $ -2.78$    & $-3.83$   &       
  1     &            &  0.36       &     0.35  \\
        & $-1.6481$  & $ -2.8635$  & $-3.9028$ &
        &            &   0.4831    &   0.4875  \\
        & $(-1.78)$  &  $(-3.25)$ & $(-4.49)$&
        &         &  $ (0.41)$        & $(0.40)$ \\

  2     & $-1.01$    & $-1.97$     & $-2.96$   &
  2     &            &             &  0.36     \\
        & $-0.9310$  & $-1.9225$   & $-2.8453$ &
        &            &             &  0.5364   \\
   & $(-1.17)$       &  $(-2.30)$ & $(-3.49)$&
        &         &               & $(0.42)$ \\
\hline
 LSE & & & & LSO & & & \\
 (D/D)   &   $n = 0$   &    1       &    2      & 
 (P/P)   &   $n = 0$   &    1       &    2      \\
\hline
$n'= 0$     &    0.085     &   0.102   & 0.105     &               
$n'= 0$     &   $-0.084$   & $-0.207$    & $-0.321$   
\\                
            & $-0.0024$   & $-0.0069$  & $-0.0094$ &
            & $-0.1928$   & $-0.2860$  & $-0.3548$ \\
	&  (0.017)   &     (0.026)     & (0.034)         &
          &  $(-0.260)$     & $(-0.400)$  & $(-0.520)$ \\ 
    1       &             &   0.145     &   0.163    &
    1       &             &  $-0.359$   & $-0.505$    \\          
            &             &  $-0.0148$ & $-0.0183$ &
            &             &  $-0.4254$ & $-0.5296$  \\
		&		  &  $(0.042)  $ & $(0.054)$&
            &             &  $(-0.610)$ & $(-0.780)$   \\
    2       &             &            &   0.198     &
    2       &             &            & $- 0.669$    \\
            &             &            & $-0.0352$  &
            &             &            & $-0.6610$\\
		&		  &            & $(0.070)     $ &
            &             &            & $(-0.980)    $ \\
 	 
\hline
\end{tabular}

\end{table}

\newpage


\begin{table}[tbp]
   \centering
   \caption{\small Best-fit interaction strength (MeV)for 
$A=24$,$\hbar\omega$=12.7 MeV, $\alpha$=0.075}
\vspace*{0.5cm}

\begin{tabular}{lllllr} 
\hline
       & Channel& $R_{1}=0.25 fm$ &  $R_{2}=0.40 fm$ & $R_{3}=0.70 
fm$ & $R_{4}= 1.414 fm$ \\ 
\hline
  1    &SE        &10790 &   -3538   &         & -10.463 \\

  2    &TE        &23209   &   -7211   &         & -10.463 \\
	
  3    &SO        &-321.4   &   2593   &         & 31.389 \\
	
  4    &TO        &421.4   &   807.1   &         & 3.488 \\
	 
  5    &TNE       &        & -1441   &  -4.8  &       \\
	
  6    &TNO        &      & -25.9   &  23.7   &       \\
	 
  7    &LSE       & -3239      & 1973   &    &       \\
	
  8    &LSO       & -66.7       & -816.7   &      &    \\
	 
\hline
\end{tabular}
\end{table}


\begin{table}[tbp]
   \centering
   \caption{\small Best-fit interaction strength (MeV)for 
$A=40$, $\hbar\omega$=11 MeV, $\alpha$=0.070}
\vspace*{0.5cm}

\begin{tabular}{lllllr} 
\hline
       & Channel& $R_{1}= 0.25 fm$ &  $R_{2} = 0.40 fm$ & $R_{3}= 
0.70 fm$ & $R_{4} = 1.414 fm$ \\ 
\hline
  1    &SE        &10427 &   -3414   &         & -10.463 \\

  2    &TE        &22035   &   -6741   &         & -10.463 \\
	
  3    &SO        &9376   &   969   &         & 31.389 \\
	
  4    &TO        &-138.9   &   798.4   &         & 3.488 \\
	 
  5    &TNE       &        & -1260   &  -19.2  &       \\
	
  6    &TNO       &      & 11.34   &  21.70   &       \\
	 
  7    &LSE       & -48365      & 5188   &    &       \\
	
  8    &LSO       & -1721       & -555   &      &    \\
	 
\hline
\end{tabular}
\end{table}


\begin{table}[tbp]
   \centering
   \caption{\small Best-fit interaction strength (MeV)for 
$A=90$,$\hbar\omega$=8.8 MeV, $\alpha$=0.060}
\vspace*{0.5cm}

\begin{tabular}{lllllr} 
\hline
       & Channel& $R_{1}= 0.25 fm$ &  $R_{2} =0. 40 fm$ & $R_{3} 
=0.70 fm$ & $R_{4} = 1.414 fm$ \\ 
\hline
  1    &SE        &8452 &   -2871   &         & -10.463 \\

  2    &TE        &17554   &   -5311   &         & -10.463 \\
	
  3    &SO        &3611   &   490   &         & 31.389 \\
	
  4    &TO        &1402   &   183   &         & 3.488 \\
	 
  5    &TNE       &        & -1204   &  -25.2  &       \\
	
  6    &TNO       &      & 61.02   &  19.02   &       \\
	 
  7    &LSE       & -40737      & 4476   &    &       \\
	
  8    &LSO       & -19239       & 1731   &      &    \\
	 
\hline
\end{tabular}
\end{table}



\newpage

\begin{thebibliography}{99}
    
\bibitem[1]{kn:Bertsch}G. Bertsch, J. Borysowicz, H. McManus and W.G. 
Love,  Nucl. Phys. {\bf A284}, 399 (1977). 

\bibitem[2]{kn:Dao}Dao T. Khoa and G. R. Satchler, Nucl. Phys. 
{\bf A668}, 3 (2000).

\bibitem[3]{kn:Ismail}M. Ismail and Kh. A. Ramadan,  J. Phys. G. 
Nucl. Part. Phys. {\bf 26}, 1621 (2000).

\bibitem[4]{kn:Delion}D.S. Delion, A. Sandulescu, S. Misicu, F. 
Carstoiu and W. Greiner, Phys. Rev {\bf C64}, 041303(R) (2001).

\bibitem[5]{kn:Fiase}J.O. Fiase, A. Hosaka, L.K.Sharma and D.P. 
Winkoun. Fundamental and Applied Aspects of Modern Physics, Luderitz 
2000 Conference Proceedings, World Scientific pub.(S.H. Connell and 
R.Tegen ed.) Namibia, 70(2000).

\bibitem[6]{kn:Satchler}G.R. Satchler and W.G. Love, Phys. Repts 
{\bf 55}, 183 (1979).

\bibitem[7]{kn:Wildenthal}B.H. Wildenthal, Prog. in Part. and Nucl. 
Phys. {\bf 11}, 5 (1984)

\bibitem[8]{kn:Fiase}J. Fiase, A. Hamoudi, J.M. Irvine and F. 
Yazici,  J. Phys. G. Nucl. Phys. {\bf 14}, 27 (1988).

\bibitem[9]{kn:Irvine}J.M. Irvine, Prog. in Part. and Nucl. Phys. 
{\bf 5}, 1 (1980).

\bibitem[10]{kn:Irvine}J.M. Irvine, G.S. Mani, V.F.E. Pucknell, M. 
Vallieres and F. Yazici, Ann. Phys. (N.Y.) {\bf 102}, 129 (1976).

\bibitem[11]{kn:Reid}R.V. Reid, Ann. Phys. (N.Y) {\bf 50}, 411 (1968).

\bibitem[12]{kn:Fiase}J. Fiase, Phys. Rev {\bf C63}, 037303-1 (2001).

\bibitem[13]{kn:Yazici}F. Yazici and J.M. Irvine,   J. Phys. G. Nucl. Phys. {\bf 13}, 615 (1987).

\bibitem[14]{kn:Anantaranman}N. Anantaraman, H. Toki and G.F Bertsch, 
Nucl. Phys. {\bf A398}, 269 (1983).
\bibitem[15]{exact} S.C. Pieper and R.B. Wiringa, 
Annu. Rev. Nucl. Part. Sci. {\bf 51}, 53 (2001);\\
R.B. Wiringa et al.,  Phys. Rev. {\bf C62}, 014001 (2000).  

\bibitem[16]{toki} H. Toki, S. Sugimoto and K. Ikeda, 
nucl-th/0110017.  

\bibitem[17]{kn:Hosaka}A. Hosaka, K. I. Kubo and H. Toki, Nucl. Phys. 
{\bf A444}, 76 (1985).


\bibitem[18]{kn:Elliot}J.P. Elliot et al., Nucl. Phys. {\bf A121}, 
241 
(1968). 

\end{thebibliography}
\end{document}